\documentclass[]{gmaart2022}

\renewcommand{\varProcName}{Proc. 34. Workshop Computational Intelligence}
\renewcommand{\varLocation}{Berlin}
\renewcommand{\varDate}{20.-21.11.2024}

\usepackage[utf8]{inputenc} 
\usepackage[T1]{fontenc} 
 
\usepackage{amsmath,amssymb,amsfonts,mathtools,amsthm} 

\usepackage[babel,autostyle]{csquotes}

\usepackage{tabularx}
\newcolumntype{L}[1]{>{\raggedright\arraybackslash}p{#1}} 
\newcolumntype{C}[1]{>{\centering\arraybackslash}p{#1}} 
\newcolumntype{R}[1]{>{\raggedleft\arraybackslash}p{#1}} 
\usepackage{booktabs}

\usepackage{paralist}   

\usepackage[noend]{algpseudocode}

\algrenewcommand\algorithmicrequire{\textbf{Voraussetzung:}}
\algrenewcommand\algorithmicensure{\textbf{Abschlussbedingung:}}

\usepackage{listings} 
\usepackage{subcaption}  

\usepackage{epstopdf}
\usepackage{xcolor}
\selectcolormodel{gray}  

\usepackage{scrhack}
\usepackage{varwidth}
\usepackage{rotating} 
\usepackage[defaultlines=2,all]{nowidow}  
\usepackage{import}
\usepackage{placeins}
\usepackage{makecell}


\usepackage[version-1-compatibility]{siunitx} 

\usepackage{amsmath,amssymb,amsfonts,mathtools,amsthm} 

\usepackage{longtable} 
\usepackage{multirow}
\usepackage{tikz}
\usepackage{pgfplots}
\usepackage{smartdiagram}
\usetikzlibrary{patterns}
\usetikzlibrary{shapes} 
\usetikzlibrary{shapes.geometric, arrows,positioning}
\usepackage{smartdiagram}
\usesmartdiagramlibrary{additions} 
\AtBeginDocument{
}
\usepackage{xcolor}
\selectcolormodel{gray}

\usepackage[acronym]{glossaries} 
\usepackage{hyphenat}
\usepackage{etoolbox}
\usepackage{printlen}
\usepackage{graphicx}
\usetikzlibrary{calc, positioning, matrix}
\usepackage{standalone}

\newcommand{\newacr}[4][]{\newacronym[
	sort={\ifthenelse{\isempty{#1}}{#2}{#1}},
	]{#2}{#3}{#4}}
\glsdisablehyper

\robustify{\bfseries}

\usepackage[capitalise]{cleveref}
\usepackage{diagbox} 
\usepackage{array}   
\usepackage{tabularx} 
\usepackage{makecell} 

\begin{document}

\hyphenpenalty=2000

\pagenumbering{roman}
\setcounter{page}{1}
\pagestyle{scrheadings}
\pagenumbering{arabic}

\setnowidow[2]
\setnoclub[2]

\renewcommand{\Title}{EAP4EMSIG - Experiment Automation Pipeline for Event-Driven Microscopy to Smart Microfluidic Single-Cells Analysis}

\renewcommand{\Authors}{
    Nils~Friederich\textsuperscript{1,2,*},
    Angelo~Jovin~Yamachui~Sitcheu\textsuperscript{1,*},
    Annika~Nassal\textsuperscript{1,2},
    Matthias~Pesch\textsuperscript{3},
    Erenus~Yildiz\textsuperscript{4},
    Maximilian~Beichter\textsuperscript{1},
    Lukas~Scholtes\textsuperscript{4},
    Bahar~Akbaba\textsuperscript{1},
    Thomas~Lautenschlager\textsuperscript{1},
    Oliver~Neumann\textsuperscript{1},
    Dietrich~Kohlheyer\textsuperscript{3},
    Hanno~Scharr\textsuperscript{4},
    Johannes~Seiffarth\textsuperscript{3,5,\#},
    Katharina~Nöh\textsuperscript{3,\#},
    Ralf~Mikut\textsuperscript{1,\#}
}
\renewcommand{\Affiliations}{
    \textsuperscript{1}
    Institute for Automation and Applied Informatics (IAI)\\
    \textsuperscript{2}
    Institute of Biological and Chemical Systems (IBCS)\\
    Karlsruhe Institute of Technology\\
    \textsuperscript{3}
    Institute of Bio- and Geosciences (IBG-1)\\
    \textsuperscript{4}
    Institute for Data Science and Machine Learning (IAS-8)\\
    Forschungszentrum Jülich GmbH\\
    \textsuperscript{5}
    Computational Systems Biology (AVT-CSB)\\
    RWTH Aachen University\\
    *Contributed equally\\
    \#Supervised equally
}

							 
\renewcommand{\AuthorsTOC}{N.~Friederich, A.~J.~Yamachui~Sitcheu, A.~Nassal, M.~Pesch, E.~Yildiz, M.~Beichter, L.~Scholtes, B.~Akbaba, T.~Lautenschlager, O.~Neumann, D.~Kohlheyer, H.~Scharr, J.~Seiffarth, K.~Nöh, R.~Mikut} 
\renewcommand{\AffiliationsTOC}{Karlsruhe Institute of Technology, Forschungszentrum Jülich GmbH, RWTH Aachen University} 

\setLanguageEnglish
							 
\setupPaper 


\newacronym{ai}{AI}{Artificial~Intelligence}

\newacronym[%
  shortplural={APs},%
  longplural={Average~Precisions}%
] {ap}{AP}{Average~Precision}

\newacronym[
  shortplural={CNNs},%
  longplural={Convolutional~Neural~Networks}%
]{cnn}{CNN}{Convolutional Neural Network}

\newacronym{cpn}{CPN}{Contour~Proposal~Network}

\newacronym{cpu}{CPU}{Central Processing Unit}

\newacronym{db}{DB}{DataBase}

\newacronym{dl}{DL}{Deep~Learning}

\newacronym{dora}{DORA}{Dataflow-Oriented~Robotic~Architecture}

\newacronym{eap}{EAP}{Experiment~Automation~Pipeline}

\newacronym{eap4emsig}{EAP4EMSIG}{Experiment~Automation~Pipeline~for~Event-Driven~Microscopy~to~Smart~Microfluidic~Single-Cells~Analysis}

\newacronym{eapdp}{EAPDP}{Experiment~Automation~Pipeline~for~Dynamic~Processes}

\newacronym[
    shortplural={GPUs},%
    longplural={Graphics Processing Units}%
]{gpu}{GPU}{Graphics Processing Unit}

\newacronym{gui}{GUI}{Graphical~User~Interface}

\newacronym{kaida}{KaIDA}{Karlsruhe~Image~Data~Annotation}

\newacronym{mae}{MAE}{Mean~Absolute~Error}

\newacronym[%
  shortplural={mAPs},%
  longplural={mean~Average~Precisions}%
] {map}{mAP}{mean~Average~Precision}

\newacronym{ml}{ML}{Machine~Learning}

\newacronym{mlci}{MLCI}{Microfluidic~Live-Cell~Imaging}

\newacronym{mlp}{MLP}{Multi-Layer~Perceptron}

\newacronym{mp}{MP}{Mixed~Precision}

\newacronym{omero}{OMERO}{Open~Microscopy~Environment~Remote~Objects}

\newacronym{poc}{PoC}{Proof~of~Concept}

\newacronym{pq}{PQ}{Panoptic~Quality}

\newacronym{pyme}{PYME}{PYthon~Microscopy~Environment}

\newacronym{ram}{RAM}{Random-Access~Memory}

\newacronym{relu}{ReLU}{Rectified~Linear~Unit}

\newacronym{ros}{ROS}{Robot~Operating~System}

\newacronym{rq}{RQ}{Recognition~Quality}

\newacronym{sota}{SOTA}{State-Of-The-Art}

\newacronym{sq}{SQ}{Segmentation~Quality}

\newacronym{tp}{TP}{True~Positive}

\newacronym{ui}{UI}{User~Interface}

\newacronym{yolo}{YOLO}{You~Only~Look~Once}


\begin{sloppy}

\section*{Abstract}
\label{section:abstract}
\gls{mlci} generates high-quality data that allows biotechnologists to study cellular growth dynamics in detail. However, obtaining these continuous data over extended periods is challenging, particularly in achieving accurate and consistent real-time event classification at the intersection of imaging and stochastic biology. To address this issue, we introduce the \gls{eap4emsig}. In particular, we present initial zero-shot results from the real-time segmentation module of our approach. Our findings indicate that among four \gls{sota} segmentation methods evaluated, Omnipose delivers the highest \gls{pq} score of 0.9336, while \gls{cpn} achieves the fastest inference time of 185 ms with the second-highest \gls{pq} score of 0.8575. Furthermore, we observed that the vision foundation model Segment Anything is unsuitable for this particular use case.


\section{Introduction}
\label{section:introduction}
\paragraph{What are microbes?}
Microbes, also known as microorganisms, are a group of tiny living organisms that are invisible to the naked eye. This group includes bacteria, archaea, fungi and protists~\cite{AmericanSocietyforMicrobiology.2022}. Microbes are present almost everywhere on Earth, from harsh environments such as hydrothermal vents to the human body, where they outnumber human cells by a factor of around 1.3~\cite{Sender.2016}. Despite their tiny size, microbes play crucial roles in various ecological and biological processes, making them essential for life on Earth~\cite{Locey.2016}.

\paragraph{Why are microbes relevant?}
Microbes are relevant for several reasons. The first is ecological balance, where microbes are essential in the nutrient cycle, decomposing organic matter and contributing to soil fertility~\cite{Sylvia.2005}. They are crucial for the carbon, nitrogen and sulfur cycles that sustain life on Earth~\cite{Li.2021}. Second, in human health, the human microbiome aids digestion, produces essential vitamins and protects against pathogenic microbes~\cite{Oliphant.2019}. Disruptions in the microbiome can lead to health issues such as infections, obesity and autoimmune diseases~\cite{Christovich.2022, Liu.2021}. Finally, in the context of industrial applications, microbes are harnessed in biotechnology, pharmaceuticals and agriculture. They are used to produce antibiotics, biofuels and fermented foods~\cite{Lancini.2013}. Microbial enzymes are also crucial in many manufacturing processes~\cite{Sanchez.2017}.

\paragraph{Why is research on microbes essential?}
Research on microbes is important due to their impact on health, industry and the environment. Understanding microbial behavior, genetics and interactions can advance all three areas. In medical science, it is crucial to study pathogens to help develop vaccines and treatments for infectious diseases~\cite{Bethe.2024}. Microbe research can potentially reveal new therapies for chronic diseases~\cite{AmericanSocietyforMicrobiology.2005}. In environmental protection, microbes can be used in bioremediation to clean up oil spills and toxic waste~\cite{Ganesan.2022}. Therefore, understanding microbial ecosystems can help conservation efforts and combat climate change. In biotechnology, microbial research can lead to the development of new applications, such as using microbes to produce valuable compounds, e.g., insulin or biodegradable plastics~\cite{Nduko.2020}.

\paragraph{Why is the segmentation of microbes relevant?}
While some biological analysis is possible at the macroscopic level, other results can only be obtained by studying organisms at the microscopic single-cell level. \gls{mlci} particularly enables an understanding of single-cell growth and growth heterogeneity due to very small volumes. For example, the effect of antibiotic concentrations on organism performance can be analyzed through such experiments. Understanding the dynamics of microbes at this single-cell level therefore requires accurate and precise automated cell segmentation, as large amounts of data acquired using automated microscopy must be analyzed to obtain relevant results. The segmented data can then be used to make statements about the organism's growth as a function of various other factors.

\paragraph{What is the challenge in microbe research?}
\gls{mlci} experiments with microbes are usually not carried out on a single colony but in parallel on thousands. To achieve this, the microfluidic device is infused with a cell suspension and cells are randomly seeded into the growth chambers, where they remain trapped. Optimally, a microbial colony grows in each chamber. In a standard growth experiment, seeded cells grow until the chamber is filled with densely packed cells, which can be 1000 ends, which marks the end of the experiment. Subsequent experiment examination requires analyzing thousands of colonies in parallel, with thousands of microbes in some cases. Each chamber must be manually assessed to determine whether it meets the experiment's objectives, even as some chambers become irrelevant as the experiment advances. This process is extremely time-consuming, costly, strenuous, monotonous and nearly impossible. Therefore, automated and intelligent processing, analysis and experiment planning are required.

\paragraph{How does this paper address this challenge?}
In this paper, we introduce the \gls{eap4emsig}, designed to automate and intelligently execute \gls{mlci} experiments, during which the human expert specifies settings, monitors progress and intervenes only to address any issues that may arise. We explain the concept of the pipeline and its eight primary modules. To achieve this, a literature review (see~\cref{section:related_work}) and an extensive description (see~\cref{section:methodology}) of each module are provided.

Since real-time data evaluation, inference and incorporation into the experimental design are central parts of our entire \gls{eap} pipeline, our work will compare initial results. We will compare the results related to the \gls{ap}~\cite{lin2014microsoft} score, \gls{pq}~\cite{kirillov2019panoptic} score and inference time of four \gls{sota} methods from the task, domain and foundation areas, using a large publicly available microbial benchmark dataset~\cite{dataMicrobsTrackingMillion,trackoneinamillion} (see~\cref{section:experiments}).
For this purpose, their zero-shot abilities and their real-time capability will be analyzed and investigated to determine which models are suitable for retraining. Additionally, we will evaluate whether using a foundation model potentially leads to better results through improved generalization.


\section{Related Work}
\label{section:related_work}

\paragraph{Experiment Automation Pipelines.}
Various \gls{eap} tools have been developed, ranging from local standalone projects~\cite{dettinger2018automated} to cloud-based tools~\cite{doi:10.1177/2472630318784506}. Some methods focus on automating the data analysis part~\cite{friederich2023ai, D3LC00327B}, others focus on automating the data acquisition part~\cite{rahmanian2022enabling}, particularly on microscope control~\cite{pinkard2021pycro, susano2021python} and event-based image acquisition~\cite{chiron2022cybersco, mahecic2022event}. However, very few generic tools for full experiment automation have been proposed due to the complexity of combining the experiments' software, hardware and biological components. One example is the \gls{pyme}\footnote{https://www.python-microscopy.org/} open-source package, which offers data acquisition, processing, exploration and visualization modules. \gls{pyme} is, however, only tailored for super-resolution techniques. Another example is Cheetah~\cite{pedone2021cheetah}, a Python library that automates real-time cybergenetic experiments. It offers limited microscope control capabilities and relies on one specific image segmentation method, i.e., U-Net~\cite{ronneberger2015u}.

Recently, the \gls{eap} tool MicroMator~\cite{fox2022enabling} has emerged, strongly aligning with our goal. Similarly to the idea of smart futuristic microscopy depicted in~\cite{carpenter2023smart} and~\cite{pinkard2022microscopes}, MicroMator supports reactive microscopy experiments. The developed open-source package is modular, extendable and adjustable for several experiments. However, they offer limited image analysis possibilities and no tool to improve the image analysis results. Moreover, the software seems not to be actively used and maintained.

In summary, while many tools exist that each contribute to a step towards fully \gls{eap}, no tool, to the best of our knowledge, provides a complete, modular and extendable pipeline that manages event-based data acquisition, analysis and reporting.

\paragraph{Segmentation.}
\label{paragraph:related_work-segmentation}
Deep learning-based segmentation methods have recently emerged as they are often faster, more accurate, and precise than traditional methods ~\cite{7881449}, given sufficient training data availability~\cite{esteva2021deep}.

A method with pixel-wise segmentation is required to obtain all the information needed for event detection in the context of microbes. To allow the extracted data to flow directly in the \gls{eap} during the experiments, this method must be fast enough, accurate and precise to enable real-time processing~\cite{lou2023cfpnet}.  Therefore, objects can be segmented, for example, with generalized methods like the \gls{sota} vision model Segment Anything~\cite{kirillov2023segany}, so-called foundation models, which attempt to recognize all objects correctly, e.g. in the context of segmentation. Although these methods can recognize many diverse objects, they may have limitations in precision for a single use-case~\cite{ji2024segment}. To overcome this problem, there are also \gls{sota} domain-specific biomedical methods like \gls{cpn}~\cite{celldetection} and StarDist~\cite{schmidt2018cell} or task-specific-models like Omnipose~\cite{cutler2022omnipose}.

With the wide variety of models available, selecting the most appropriate one for a given task remains a significant challenge. Currently, this selection is still performed manually. However, solutions that aim to automate this selection process are being proposed. ~\cite{figerprint_lena_maier_heine, fingerprint_marcel, yamachui_sitcheu_mlops_2023} investigate image similarity metrics to select the best model for a given task.

\paragraph{Experiment Database.}
\label{paragraph:related_work-exp_db}
\gls{mlci} experiments produce vast amounts of data. This data and associated metadata must be stored and managed for subsequent analysis and reporting. In the context of \gls{eap4emsig}, the data management tool must support the FAIR data management principles as depicted in~\cite{wilkinson2016fair}.

For our work, the most suitable tool is \gls{omero}~\cite{allan2012omero}, an open-source tool for managing, analyzing and visualizing microscopy images and their metadata. It offers a centralized, secure and scalable solution for handling diverse imaging data types, facilitating collaboration and data sharing among entities. Compared to other \gls{sota} data management tools, \gls{omero} provides advanced visualization tools and supports integration with other image analysis software~\cite{schmidt2022research}.

\paragraph{Semi-Automated Data Annotations.}
\label{paragraph:related_work-ss_data_anno}
To train the segmentation methods, corresponding training data are crucial. Annotating this data is typically time-consuming, so semi-automated segmentation tools like KaIDA~\cite{schilling2022kaida} or ObiWan-Microbi~\cite{obiwanMicroby_microbseg} are often employed in biomedical use cases~\cite{kirillov2023segany,xiao2024florence,scherr2022microbeseg}. This process involves training a network with a small amount of manually annotated datasets and manually refining the network's predictions by a human annotator on new unannotated datasets.

\paragraph{AI-ready Image with Ground Truth Cell Simulation.}
\label{paragraph:related_work-cell_simulation}
A significant challenge in applying \gls{dl} techniques is the need for labeled data for training and validation. Particularly in cell instance segmentation tasks, pixel-exact masks that accurately distinguish individual cells from the background are essential. Due to the high costs needed to generate such labeled data~\cite{10.1093/femsre/fuaa062}, cell simulators have been developed~\cite{lehmussola2007computational, svoboda2016mitogen}. Their aim is to create realistic, controlled and reproducible cellular models that accurately reflect biological processes. For bacterial microcolony ground truth generation, particularly in the context of phase contrast microscopy, the cell simulator CellSium~\cite{sachs2022cellsium} emerges as the suitable tool in this work. It is an agent-based, highly customizable and versatile simulator that produces data for different use cases.

\paragraph{Module Interaction.}
\label{paragraph:related_work-module_interaction}
Given the complexity of integrating software, hardware and biological components in laboratory experiments, a suitable architecture is required. This architecture must be robust, understandable, modular and most importantly extendable due to the uniqueness of each laboratory experiment. For our \gls{eap}, we currently use \gls{ros}~\cite{ros2}, an open-source framework primarily for developing robot software.

\gls{ros} provides a modular building architecture based on the central notion of nodes. Each node represents a functional unit and can exchange messages with another node, particularly in an event-based manner. Hence, \gls{ros} is very suitable for real-time tasks in various fields, such as in~\cite{he2023design}. Nevertheless, due to the high complexity of installing and maintaining \gls{ros} as well as its dependency bugs~\cite{fischer2020forgotten}, just very few approaches use it. The closest to ours is Archemist~\cite{fakhruldeen2022archemist}, an experiment-automating system for chemistry laboratories.

An alternative to \gls{ros} which is currently being investigated for our \gls{eap} is \gls{dora}\footnote{https://dora-rs.ai/}, a framework designed to ease and simplify the development of AI-based robotic applications. To the best of our knowledge, \gls{dora} is very new and has not been used for experiment automation tasks yet. It provides low-latency, composable and distributed dataflow capabilities. Applications are organized as directed graphs, often referred to as pipelines. Although it is much faster than \gls{ros}, it is still unstable and has a rather smaller community.

\section{Methodology}
\label{section:methodology}
To fill the noted gaps, we propose a new \gls{eap} approach, which is fully described module-by-module in this section. As shown in~\cref{fig:exp_auto_pipeline}, our system consists of eight modules arranged in a cyclical process. For image acquisition, the system utilizes \gls{sota} research microscope setups and low-cost 3D-printed microscope systems in the first \gls{eap} module. Second, the real-time image processing is executed on incoming images, generating single-cell instance segmentation predictions. The generated data and metadata are saved and managed in an instance of the third module's \gls{omero} \gls{db}. This instance is also used to manage ground truth data generated with the cell simulator module CellSium and the ObiWan-Microbi semi-annotation module as the fourth and fifth modules. Sixth, the real-time data analysis module relies on this data to provide feedback via a dashboard and detect events. Based on these, the real-time experiment planner, as the seventh module, schedules the experiment continuously and sends the next steps to the microscope control module, the eighth module, which forwards these instructions back to the image acquisition module.
    \begin{figure}[tb]
        \centering
        \includegraphics[width=0.95\linewidth]{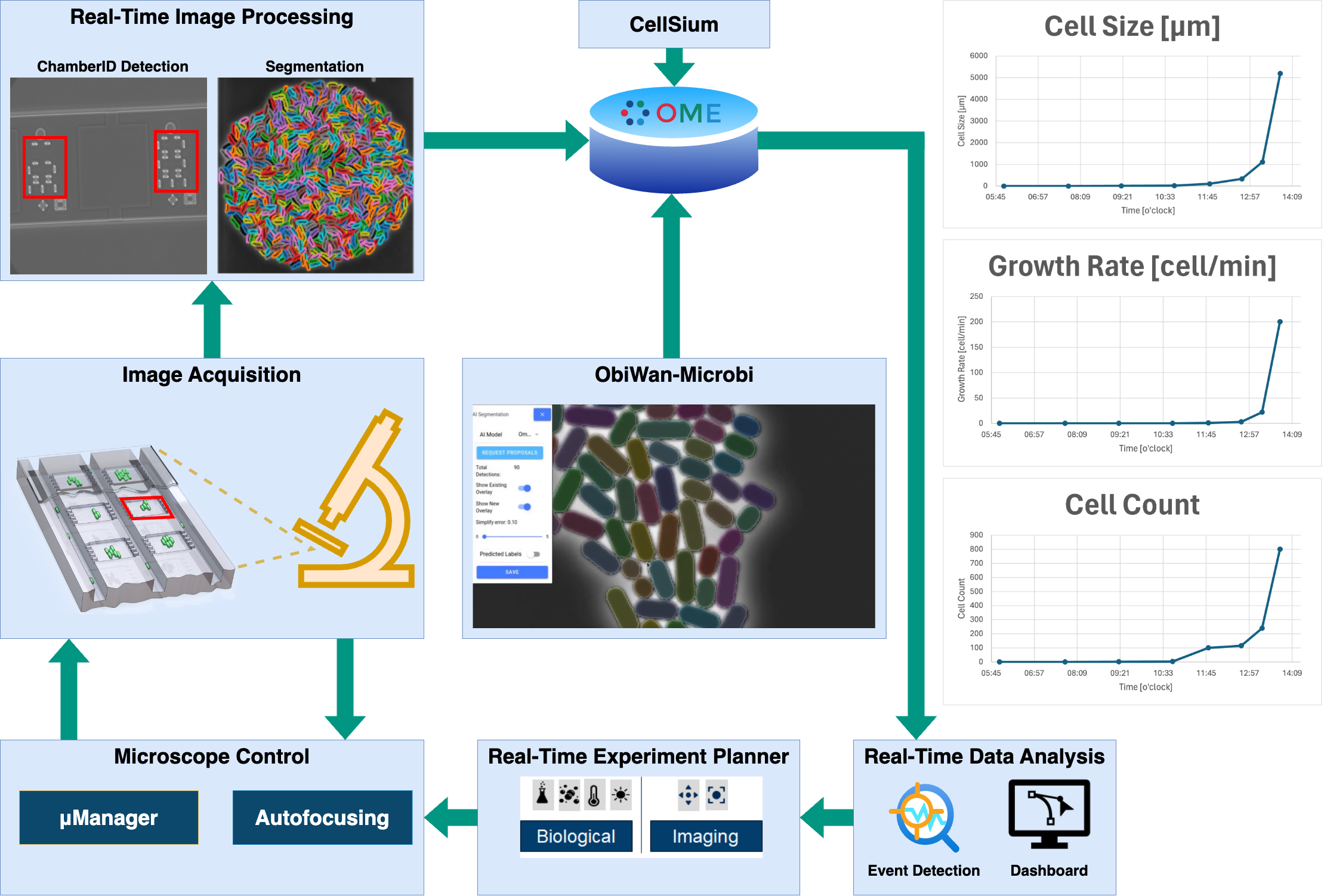}
        \caption{EAP4EMSIG visualization. The pipeline consists of eight modules, represented by the light blue boxes and the \gls{omero} database, arranged in a cyclical process. The microbial images in the figure come from dataset~\cite{dataMicrobsTrackingMillion}. The images from the experiment chip are from an internal dataset.}
        \label{fig:exp_auto_pipeline}
    \end{figure}
The modules are described individually in the upcoming sections.

\subsection{Microscope Control}
\label{subsection:methodology-microscope_control}
\paragraph{µManager.} Automatic control of the microscope is essential for experiment automation. To make our \gls{eap} as independent as possible from microscope manufacturers and thus enable easy transfer to new laboratories and microscopes, µManager~\cite{edelstein2014advanced} is used. µManager is an open-source software package that controls microscopes and associated hardware components such as cameras, stages and shutters. It provides a powerful, flexible and cost-effective solution for automated microscopy. In this work, the implementation is done in Python with the help of Pymmcore(-Plus)\footnote{https://github.com/micro-manager/pymmcore} with a Nikon T1-based setup.

\paragraph{Autofocusing.} One specific challenge we address here is the autofocusing of the microscope. We treat autofocusing as a regression problem, where a simple \gls{mlp} is used to predict the relationship between the extracted input features from microscopy images and the continuous target variable, which is the distance to the optimal focus frame (among all z-stacks). After predicting the focus offset and direction, our system employs a closed-loop control mechanism to communicate the predicted adjustments to the microscope control. The focus is then iteratively adjusted until the optimal focus is reached.

\subsection{Image Acquisition}
\label{subsection:methodology-image_acquisition}
This module handles image acquisition primarily in two different ways:
    \begin{enumerate}
        \item The most common process is using real research microscopes for the experiments and generating high-quality images. This is still by far the most used approach, particularly due to the ability of such microscopes to provide direct, high-resolution and high-fidelity images of biological samples.
        \item The low-cost alternative to such expensive tools are 3D-printed microscopes, which are emerging as cost-effective and accessible tools, especially in educational settings and low-resource environments~\cite{Collins:20}. However, they generally do not match the resolution and functionality of high-end commercial microscopes.
    \end{enumerate} 
The acquired image data and metadata are then collected and saved according to standardized protocols in a \gls{omero} \gls{db}. Standardization offers the possibility of a uniform mask for querying different information for all modules (including future ones). Furthermore, the data can be distributed, stored and accessed worldwide.

\subsection{Real-Time Image Processing}
\label{subsection:methodology-rt_img_proc}
The image data acquired in the previous step (see~\cref{subsection:methodology-image_acquisition}) is processed as shown in~\cref{fig:exp_auto_pipeline}. On the one hand, the region of interest, that is the growth chamber, is extracted by removing any microfluidic structures from the images. On the other hand, the content of the chamber is segmented using a suitable method. In this work, we focus on \gls{sota} \gls{dl} segmentation methods (see~\cref{paragraph:related_work-segmentation}), which are either task-specific, domain-specific or foundation models and therefore allow us to address various segmentation tasks effectively. We also investigate the data processing speed of these methods. This is important because the classification of the events and, therefore, the decision of the experiment planner (see~\cref{subsection:methodology-rt_exp_planner}) is highly based on the segmentation results.

\subsection{OMERO Database}
\label{subsection:methodology-omero_db}

As mentioned in ~\cref{paragraph:related_work-segmentation}, we use \gls{omero} to manage not only the images and the associated metadata in a centralized and standardized manner but also the results of downstream analyses such as chamber detection and extraction, segmentation and cell analysis. All modules of the \gls{eap} (see~\cref{fig:exp_auto_pipeline}) can retrieve information via a standardized interface. In addition, this makes it possible for the human expert to easily and comprehensibly document his experiments, including access to the post-processing and -analysis results.

\subsection{ObiWan-Microbi}
\label{subsection:methodology-obiWan_microbi}
For intra- and inter-cell analysis to be possible, the best feasible extraction of objects through segmentation (see~\cref{subsection:methodology-rt_img_proc}) is required. One challenge in our context is the large number of densely packed cells that need to be segmented. To date, there are no labeled datasets that accurately represent a comparable use case, which would facilitate transfer learning or the training of supervised segmentation methods. Since manual labeling alone would be too long and too inefficient, the semi-automated annotation tool ObiWan-Microbi is used in this work. This tool allows the prediction and correction of labels and subsequent retraining of the used \gls{dl} segmentation models. An example of a dataset created with this is \cite{dataMicrobsTrackingMillion}, which will be used to evaluate the segmentation methods in~\cref{section:experiments}.

\subsection{CellSium}
\label{subsection:methodology-cellsium}
However, even the creation of labels using semi-automated methods such as ObiWan-Microbi (see~\cref{subsection:methodology-obiWan_microbi}) costs a lot of human time and therefore money, especially in the first iteration because the segmentation methods still provide right-angled pre-segmentations. An alternative here is to have an initial basis for the segmentation methods by using automatically generated images with associated labels, e.g., from simulations. The simulator CellSium is used in our work. CellSium enables the generation of microbe images in different growth stages and also in the density and frequency required in our context. Even if these images cannot represent the full diversity of real images, combining data augmentation methods can lead to first stable results as shown in ~\cite{sachs2022cellsium}, where only slight adjustments have to be made in ObiWan-Microbi.

\subsection{Real-Time Data Analysis}
\label{subsection:methodology-rt_data_analysis}

\subsubsection{Dashboard}
\label{subsubsection:methodology-rt_data_analysis_dashboard}
Once the microbes have been segmented, single-cell data such as average cell size and growth rate are computed and saved.  This data is visualized and, most importantly, leveraged by the human expert to navigate through the experiment. For this purpose, a customized dashboard is being developed. The dashboard provides real-time insights into ongoing experiments such as cell count, growth rate and average cell size per chamber. The dashboard integrates various functionalities to facilitate the monitoring and analysis of the experiment. Due to its modular architecture, which facilitates the seamless integration of new features and functionalities without disrupting the existing codebase, our dashboard is highly extendable and can be tailored for other use cases.

\subsubsection{Event Detection}
\label{subsubsection:methodology-rt_data_event_detection}
The data and metadata gathered from the real-time data analysis and image processing enable us to detect different events in hundreds of parallel experiments and resolve their temporal evolution. In our case, we have two classes of events. On the one hand, technical events that are related to the devices themselves, e.g., loss of focus or chamber defects. On the other hand, we have biological events that display the behavior of microbes, such as growth rate or cell death. The detection is performed based on rules provided by the domain expert.

\subsection{Real-Time Experiment Planner}
\label{subsection:methodology-rt_exp_planner}
A central part of the \gls{eap} pipeline is the intelligent experiment planner. The next n chambers to be explored are determined based on the last chamber recorded, including the resulting data analysis. The determination is made according to the defined experiment objectives of the human domain expert.


\section{Experiments}
\label{section:experiments}
In this section, preliminary experiments and results of our approach, particularly for real-time image segmentation, are presented and discussed.
Four segmentation algorithms are compared on an Ubuntu 22.04-based workstation with an Intel Core i9-13900 \gls{cpu}, a RTX3090 \gls{gpu} and a 64~GB \gls{ram}. This system was chosen as low-performance because the hardware components represent an affordable system for users interested in such use cases. The measured inference time can be considered realistic for a lower boundary. An improved hardware configuration can achieve an additional performance boost here. We define 100~ms as the maximal limit for the real-time inference time. This is because initial tests of the microscope control program have shown that it is perfectly sufficient for the \gls{eap4emsig}, including autofocusing.

\subsection{Dataset, Metrics and Implementation}
\label{subsection:experiments-dataset_matrics}
The benchmark dataset~\cite{dataMicrobsTrackingMillion} is used to evaluate the methods. The dataset contains images of \textit{Corynebacterium glutamicum} microbes and represents a typical experiment in cell culture. The dataset includes video sequences of the development of the microbes with 800 images each and consists of ground truth instance segmentation mask and tracking paths. For the context of this work, we use all 5 × 800 images purely to evaluate the segmentation performances.

To evaluate the segmentation accuracy, the metrics \gls{ap}, including \gls{ap}@0.50 and \gls{ap}@0.75 and \gls{pq},  comprising \gls{sq} and \gls{rq}, are calculated for all four methods mentioned in~\cref{section:related_work}
 (see~\cref{tab:experiments-od_methods-repro-results}) using their respective official implementations.

Since the \gls{ap}-based metric requires the confidence score for calculation, evaluating this metric on Omnipose was impossible. Omnipose does not directly return uncertainties for predicted masks and no official instructions on how to extract these are known until the publication of the work.

\subsection{Real-Time Image Processing: Segmentation}
\label{subsection:experiments-rt_img_proc_seg}
The evaluation results of the four methods are shown in~\cref{tab:experiments-od_methods-repro-results}. In addition, an example image from the dataset (see~\cref{fig:results_org}) and the respective segmentation results (see~\cref{fig:results_omnipose} to~\cref{fig:results_sam}) are displayed for a medium population density with approx. 400 microbes (see~\cref{fig:results_seg}).

    \begin{figure}[tb]
        \centering
        \begin{subfigure}[b]{0.3\textwidth}
            \includegraphics[width=\linewidth]{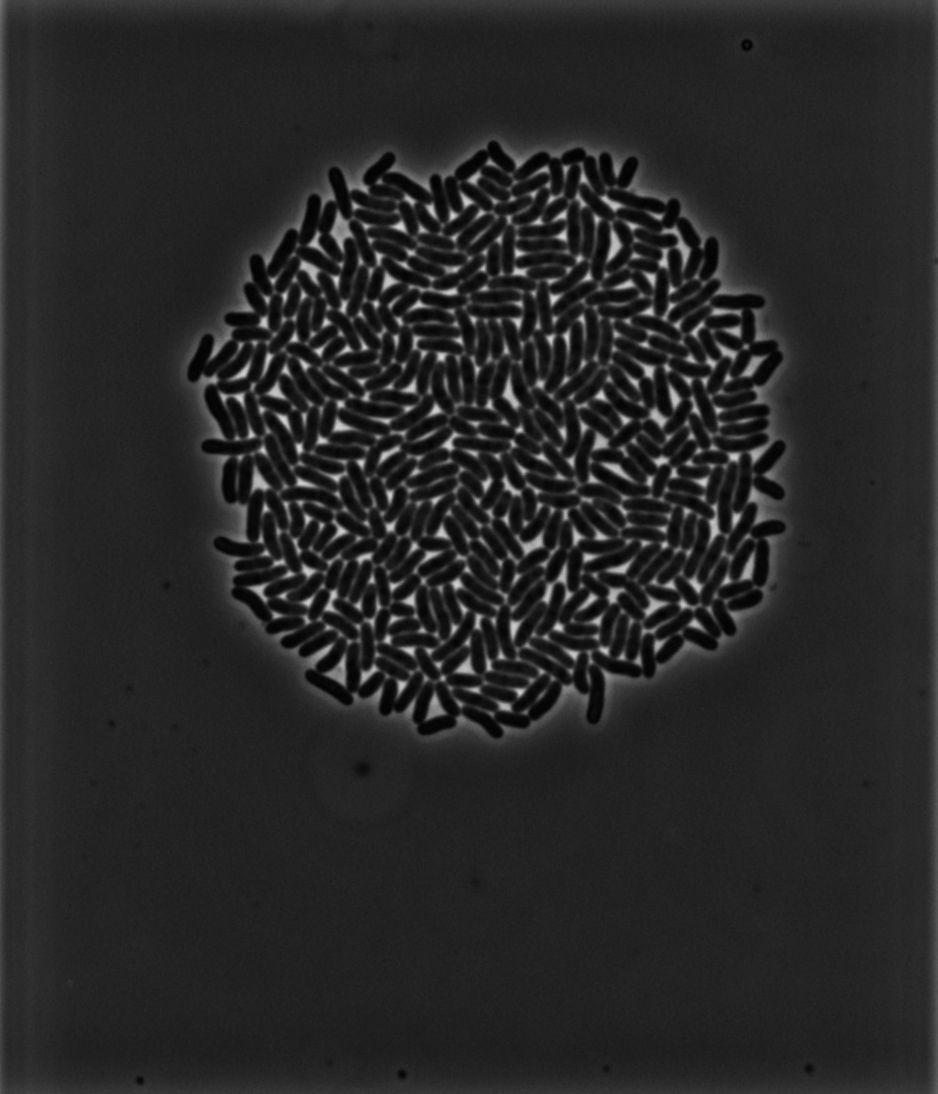}
            \caption{Original}
            \label{fig:results_org}
        \end{subfigure}
        \begin{subfigure}[b]{0.3\textwidth}
            \includegraphics[width=\linewidth]{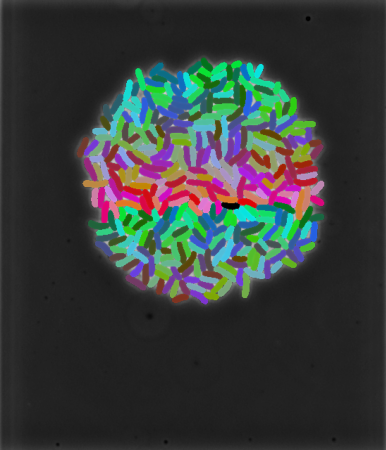}
            \caption{Omnipose~\cite{cutler2022omnipose}}
            \label{fig:results_omnipose}
        \end{subfigure}
        \begin{subfigure}[b]{0.3\textwidth}
            \includegraphics[width=\linewidth]{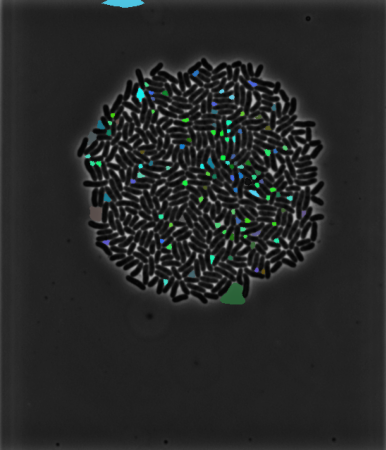}
            \caption{StarDist~\cite{schmidt2018cell}}
            \label{fig:results_stardist}
        \end{subfigure}
        \begin{subfigure}[b]{0.3\textwidth}
            \includegraphics[width=\linewidth]{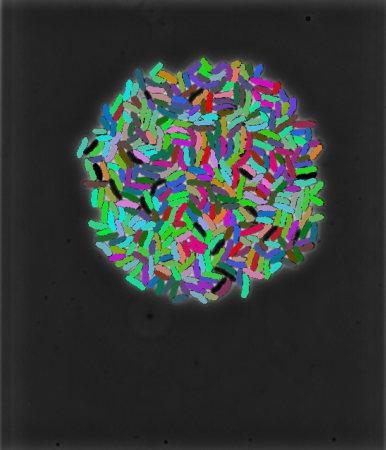}
            \caption{CPN~\cite{celldetection}}
            \label{fig:results_cpn}
        \end{subfigure}
        \begin{subfigure}[b]{0.3\textwidth}
            \includegraphics[width=\linewidth]{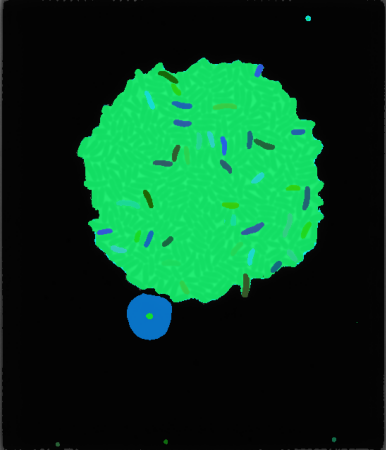}
            \caption{SAM~\cite{kirillov2023segany}}
            \label{fig:results_sam}
        \end{subfigure}

        \caption{Comparison of zero-shot instance segmentation predictions for~\cite{dataMicrobsTrackingMillion}. The original image is shown in~\cref{fig:results_org} and the predictions are shown in~\cref{fig:results_omnipose} to~\cref{fig:results_sam}.}
        \label{fig:results_seg}
    \end{figure}
   
From the results in~\cref{tab:experiments-od_methods-repro-results}, Omnipose is the best model concerning the scores \gls{pq}, \gls{sq} and \gls{rq}. However, \gls{cpn} with a \gls{pq} score difference of 0.0761, i.e., domain-specific model, is not that far away and is still 86 ms faster than Omnipose.
In detail, \gls{cpn} has a slightly lower \gls{rq} score of 0.0526, but the difference in \gls{pq} score is primarily due to the notably worse \gls{sq}. The \gls{sq} score can be seen by directly comparing~\cref{fig:results_omnipose} and~\cref{fig:results_cpn}. While Omnipose segmented the objects cleanly, including at the edges, \gls{cpn} struggled. Additionally, some parts are often no longer properly recognized in the curved microbes towards the end.

\begin{table}[tb]
    \centering
    \setlength{\tabcolsep}{2pt} 
    \begin{tabularx}{0.9125\textwidth}{|l!{\vrule width 1.25pt}c|c|c|c|}
        \hline
        \diagbox{Metric}{Method}  & Omnipose~\cite{cutler2022omnipose} & StarDist~\cite{schmidt2018cell} & CPN~\cite{celldetection} & SAM-H~\cite{kirillov2023segany} \\
        \Xhline{1.25pt}
        AP$\uparrow$ & - & 0.0000 & \textbf{0.6232} & 0.0347 \\
        AP@0.5$\uparrow$ & - & 0.0000 & \textbf{0.9551} & 0.0476 \\
        AP@0.75$\uparrow$ & - & 0.0000 & \textbf{0.8170} & 0.0470 \\
        \hline
        PQ$\uparrow$ & \textbf{0.9336} & 0.3629 & 0.8575 & 0.0626 \\
        PQ-SQ$\uparrow$ & \textbf{0.9395} & 0.7287 & 0.8779 & 0.8416 \\
        PQ-RQ$\uparrow$ & \textbf{0.9935} & 0.4093 & 0.9763 & 0.0736 \\
        \hline
        $\varnothing$Inf.~[ms]$\downarrow$ & 271 & 7686 & \textbf{185} & 1994 \\
        \hline
        
    \end{tabularx}
    \caption{Average Precision (AP) results, Panoptic Quality (PQ) results comprising Segmentation Quality (SQ) score and Recognition Quality (RQ) score as well as inference times (Inf.)  evaluated on the benchmark dataset~\cite{dataMicrobsTrackingMillion}.\protect\footnotemark}
    \label{tab:experiments-od_methods-repro-results}

\end{table}

Nevertheless, \gls{cpn}'s performance is quite remarkable because in \gls{cpn}'s training datasets, there were no such long, rod-shaped objects or similar microbial colonies in contrast to the task domain model Omnipose. As a domain model, it is also remarkable that \gls{cpn} recognizes the object instances with a similarly good \gls{rq} score (only 0.0172 difference). Both the number of objects and the difficult boundaries between the objects are not new for \gls{cpn} and also occur, for example, in \textit{NeurIPS 22 Cell Segmentation Competition}\footnote[5]{https://neurips22-cellseg.grand-challenge.org/} dataset as one of the pre-training datasets. However, in combination with the microbes, this is a noteworthy generalization achievement.

The second domain model StarDist, on the other hand, with a \gls{pq} score of 0.3629 and an \gls{ap} score of 0, has not yielded sufficient results and is clearly worse than \gls{cpn}.
The vision foundation model Segment Anything, the fourth model, is also not convincing with a \gls{pq} score of 0.0626. However, it is worth noting that the \gls{sq} score is only slightly lower than that of \gls{cpn}, indicating that the objects recognized as \gls{tp} were segmented well. However, upon examining \gls{rq}, it appears that the number of \gls{tp} is likely very low. Segment Anything's problem can also be seen in~\cref{fig:results_sam}. There, almost the entire cluster of microbes is predicted as one object. Although there are many objects in dense clusters in the dataset SA-1B~\cite{kirillov2023segany} on which the model was trained and also with a comparable number, it is quite possible that this transferred knowledge cannot be applied to the shape, which in turn raises doubts about the “Any” segmentation.

\footnotetext{When calculating the metric, falsely detected backgrounds were not removed and evaluated during the \gls{ap} calculation as false positives. The models were used according to the basic configurations for fair comparison. The values in bold are the best across all methods, provided there were results. To ensure a fair comparison, we define inference time as the duration from inputting the image to receiving the model's prediction as an instance mask with confidence scores. This includes any post-processing needed by certain methods, such as converting predicted contours to a pixel-wise mask for the \gls{eap4emsig} pipeline. The inference time is measured with all models using FP32 precision.}

\section{Conclusion}
\label{section:conclusion}

This paper presents the \gls{eap4emsig} - a novel pipeline for experiment automation for thousands of microbe colonies on microfluidic chips.
For this purpose, the methodological concept of each of the eight pipeline modules was introduced, explained and distinguished from existing alternatives. Preliminary development steps of the pipeline were presented, particularly for the real-time image segmentation module. To this end, four \gls{sota} methods were compared and evaluated quantitatively and qualitatively in the paper. \gls{cpn} and Omnipose proved to be particularly powerful. Omnipose, trained task-specifically for bacteria segmentation, is 86 ms slower at inference than \gls{cpn} but has a slightly better recognition quality and a noticeably higher segmentation quality. However, because \gls{cpn} was not trained explicitly for bacterial segmentation, but on very diverse biomedical cells such as blood cells or nuclei of different cell types, among others, future work would investigate retraining different methods to get the best model for real-time segmentation in the \gls{eap4emsig}. Future work will also investigate increasing segmentation speed to achieve the minimum 100~ms required for our task, such as converting models to special inference formats like TensorRT\footnote{https://github.com/NVIDIA/TensorRT} or transforming the trained models to a lower precision (e.g. FP16, Int8).

Even though only initial results for the real-time image processing module were shown in this work, the other modules are also being developed. So far, the pipeline as a whole has not yet been tested, but the modules themselves are already in advanced development and, in some cases, ready for use, such as CellSium or ObiWan-Microbi. The next steps are to combine the modules and test the \gls{eap4emsig} as a whole.


\section*{Acknowledgments}
\label{section:acknowledges}

This work was supported by the President's Initiative and Networking Funds of the Helmholtz Association of German Research Centres [Grant EMSIG ZT-I-PF-04-44]. The Helmholtz Association funds this project under the "Helmholtz Imaging Platform", the authors N.~Friederich, A.~J.~Yamachui~Sitcheu and R.~Mikut under the program "Natural, Artificial and Cognitive Information Processing (NACIP)", the authors N.~Friederich and A.~J.~Yamachui~Sitcheu through the graduate school "Helmholtz Information \& Data Science School for Health (HIDSS4Health)" and the author Johannes Seiffarth through the graduate school "Helmholtz School for Data Science in Life, Earth and Energy (HDS-LEE)".

The authors have accepted responsibility for the entire content of this manuscript
and approved its submission. We describe here the individual contributions of N.~Friederich (NF), A.~J.~Yamachui~Sitcheu (AJYS), A.~Nassal (AN), M.~Pesch (MP), E.~Yildiz (EY), M.~Beichter (MB), L.~Scholtes (LS), B.~Akbaba (BA), T.~Lautenschlager (TL), O.~Neumann (ON), D.~Kohlheyer (DK), H.~Scharr (HS), J.~Seiffarth (JS), K.~Nöh (KN), R.~Mikut (RM): Conceptualization: NF, AJYS, JS, RM; Methodology: NF, AJYS, EY, JS, DK, HS, KN, RM; Software: NF, AN, MB; Investigation: NF, AJYS, JS; Resources: JS, KN; Writing – Original Draft: NF, AJYS, MP, MB, ON, EY, JS; Writing – Review \& Editing: NF, AJYS, AN, MP, EY, MB, LS, BA, TL, ON, NK, HS, JS, KN, RM; Supervision: DK, HS, KN, RM; Project administration: DK, HS, KN, RM; Funding Acquisition: DK, HS, EY, KN, RM.


\end{sloppy}

\addtocontents{toc}{\protect\newpage}




\begin{thebibliography}{99}	
    \bibitem{allan2012omero}
    C.~Allan, J.-M. Burel, J.~Moore, C.~Blackburn, M.~Linkert, S.~Loynton,
      D.~MacDonald, W.~J. Moore, C.~Neves, A.~Patterson, et~al.
    \newblock {OMERO: flexible, model-driven data management for experimental
      biology}.
    \newblock {\em Nature Methods}, 9(3):245--253, 2012.
    
    \bibitem{Bethe.2024}
    U.~Bethe, Z.~D. Pana, C.~Drosten, H.~Goossens, F.~K{\"o}nig, A.~Marchant,
      G.~Molenberghs, M.~Posch, P.~{van Damme}, and O.~A. Cornely.
    \newblock Innovative approaches for vaccine trials as a key component of
      pandemic preparedness - a white paper.
    \newblock {\em Infection}, 2024.
    
    \bibitem{AmericanSocietyforMicrobiology.2005}
    K.~M. Carbone, R.~B. Luftig, and M.~Buckley.
    \newblock {\em {Microbial Triggers of Chronic Human Illness}}.
    \newblock Am Soc Microbiol, 2005.
    
    \bibitem{carpenter2023smart}
    A.~E. Carpenter, B.~A. Cimini, and K.~W. Eliceiri.
    \newblock Smart microscopes of the future.
    \newblock {\em Nature Methods}, 20(7):962--964, 2023.
    
    \bibitem{AmericanSocietyforMicrobiology.2022}
    A.~Casadevall.
    \newblock {Microbes and Climate Change - Science, People \& Impacts}.
    \newblock 2022.
    
    \bibitem{chiron2022cybersco}
    L.~Chiron, M.~Le~Bec, C.~Cordier, S.~Pouzet, D.~Milunov, A.~Banderas, J.-M.
      Di~Meglio, B.~Sorre, and P.~Hersen.
    \newblock {CyberSco. Py an open-source software for event-based, conditional
      microscopy}.
    \newblock {\em Scientific Reports}, 12(1):11579, 2022.
    
    \bibitem{Christovich.2022}
    A.~Christovich and X.~M. Luo.
    \newblock Gut microbiota, leaky gut, and autoimmune diseases.
    \newblock {\em Frontiers in Immunology}, 13:946248, 2022.
    
    \bibitem{Collins:20}
    J.~T. Collins, J.~Knapper, J.~Stirling, J.~Mduda, C.~Mkindi, V.~Mayagaya, G.~A.
      Mwakajinga, P.~T. Nyakyi, V.~L. Sanga, D.~Carbery, L.~White, S.~Dale, Z.~J.
      Lim, J.~J. Baumberg, P.~Cicuta, S.~McDermott, B.~Vodenicharski, and
      R.~Bowman.
    \newblock {Robotic microscopy for everyone: the OpenFlexure microscope}.
    \newblock {\em Biomed. Opt. Express}, 11(5):2447--2460, 2020.
    
    \bibitem{cutler2022omnipose}
    K.~J. Cutler, C.~Stringer, T.~W. Lo, L.~Rappez, N.~Stroustrup,
      S.~Brook~Peterson, P.~A. Wiggins, and J.~D. Mougous.
    \newblock Omnipose: a high-precision morphology-independent solution for
      bacterial cell segmentation.
    \newblock {\em Nature Methods}, 19(11):1438--1448, 2022.
    
    \bibitem{dettinger2018automated}
    P.~Dettinger, T.~Frank, M.~Etzrodt, N.~Ahmed, A.~Reimann, C.~Trenzinger,
      D.~Loeffler, K.~D. Kokkaliaris, T.~Schroeder, and S.~Tay.
    \newblock Automated microfluidic system for dynamic stimulation and tracking of
      single cells.
    \newblock {\em Analytical Chemistry}, 90(18):10695--10700, 2018.
    
    \bibitem{edelstein2014advanced}
    A.~D. Edelstein, M.~A. Tsuchida, N.~Amodaj, H.~Pinkard, R.~D. Vale, and
      N.~Stuurman.
    \newblock {Advanced methods of microscope control using $\mu$Manager software}.
    \newblock {\em Journal of Biological Methods}, 1(2), 2014.
    
    \bibitem{esteva2021deep}
    A.~Esteva, K.~Chou, S.~Yeung, N.~Naik, A.~Madani, A.~Mottaghi, Y.~Liu,
      E.~Topol, J.~Dean, and R.~Socher.
    \newblock Deep learning-enabled medical computer vision.
    \newblock {\em npj Digital Medicine}, 4(1):5, 2021.
    
    \bibitem{fakhruldeen2022archemist}
    H.~Fakhruldeen, G.~Pizzuto, J.~Glowacki, and A.~I. Cooper.
    \newblock Archemist: Autonomous robotic chemistry system architecture.
    \newblock In {\em {2022 International Conference on Robotics and Automation
      (ICRA)}}, pages 6013--6019. IEEE, 2022.
    
    \bibitem{7881449}
    M.~S. Fasihi and W.~B. Mikhael.
    \newblock Overview of current biomedical image segmentation methods.
    \newblock In {\em {2016 International Conference on Computational Science and
      Computational Intelligence (CSCI)}}, pages 803--808, 2016.
    
    \bibitem{fischer2020forgotten}
    A.~Fischer-Nielsen, Z.~Fu, T.~Su, and A.~W{\k{a}}sowski.
    \newblock The forgotten case of the dependency bugs: on the example of the
      robot operating system.
    \newblock In {\em Proceedings of the ACM/IEEE 42nd International Conference on
      Software Engineering: Software Engineering in Practice}, pages 21--30, 2020.
    
    \bibitem{fox2022enabling}
    Z.~R. Fox, S.~Fletcher, A.~Fraisse, C.~Aditya, S.~Sosa-Carrillo, J.~Petit,
      S.~Gilles, F.~Bertaux, J.~Ruess, and G.~Batt.
    \newblock {Enabling reactive microscopy with MicroMator}.
    \newblock {\em Nature Communications}, 13(1):2199, 2022.
    
    \bibitem{friederich2023ai}
    N.~Friederich, A.~J. Yamachui~Sitcheu, O.~Neumann,
      S.~Eroglu-Kay{\i}k{\c{c}}{\i}, R.~Prizak, L.~Hilbert, and R.~Mikut.
    \newblock {AI-based automated active learning for discovery of hidden dynamic
      processes: A use case in light microscopy}.
    \newblock In {\em Proceedings-33. Workshop Computational Intelligence: Berlin,
      23.-24. November 2023}, volume~23, page~31. KIT Scientific Publishing, 2023.
    
    \bibitem{Ganesan.2022}
    M.~Ganesan, R.~Mani, S.~Sai, G.~Kasivelu, M.~K. Awasthi, R.~Rajagopal, N.~I.
      {Wan Azelee}, P.~K. Selvi, S.~W. Chang, and B.~Ravindran.
    \newblock Bioremediation by oil degrading marine bacteria: An overview of
      supplements and pathways in key processes.
    \newblock {\em Chemosphere}, 303(Pt 1):134956, 2022.
    
    \bibitem{figerprint_lena_maier_heine}
    P.~Godau and L.~Maier-Hein.
    \newblock {Task Fingerprinting for Meta Learning in Biomedical Image Analysis}.
    \newblock In {\em {Medical Image Computing and Computer-Assisted Intervention
      -- MICCAI 2021}}, pages 436--446. Springer, 2021.
    
    \bibitem{he2023design}
    F.~He and L.~Zhang.
    \newblock Design of indoor security robot based on robot operating system.
    \newblock {\em Journal of Computer and Communications}, 11(5):93--107, 2023.
    
    \bibitem{10.1093/femsre/fuaa062}
    H.~Jeckel and K.~Drescher.
    \newblock Advances and opportunities in image analysis of bacterial cells and
      communities.
    \newblock {\em FEMS Microbiology Reviews}, 45, 2020.
    
    \bibitem{ji2024segment}
    W.~Ji, J.~Li, Q.~Bi, T.~Liu, W.~Li, and L.~Cheng.
    \newblock {Segment Anything Is Not Always Perfect: An Investigation of SAM on
      Different Real-world Applications}.
    \newblock {\em {Machine Intelligence Research}}, 21(4):617--630, Aug 2024.
    
    \bibitem{kirillov2019panoptic}
    A.~Kirillov, K.~He, R.~Girshick, C.~Rother, and P.~Doll{\'a}r.
    \newblock Panoptic segmentation.
    \newblock In {\em {Proceedings of the IEEE/CVF Conference on Computer Vision
      and Pattern Recognition}}, pages 9404--9413, 2019.
    
    \bibitem{kirillov2023segany}
    A.~Kirillov, E.~Mintun, N.~Ravi, H.~Mao, C.~Rolland, L.~Gustafson, T.~Xiao,
      S.~Whitehead, A.~C. Berg, W.-Y. Lo, P.~Doll{\'a}r, and R.~Girshick.
    \newblock {Segment Anything}.
    \newblock {\em arXiv:2304.02643}, 2023.
    
    \bibitem{Lancini.2013}
    G.~Lancini and A.~L. Demain.
    \newblock Bacterial pharmaceutical products.
    \newblock In E.~Rosenberg, E.~F. DeLong, S.~Lory, E.~Stackebrandt, and
      F.~Thompson, editors, {\em The Prokaryotes}, pages 257--280. {Springer Berlin
      Heidelberg}, Berlin, Heidelberg, 2013.
    
    \bibitem{lehmussola2007computational}
    A.~Lehmussola, P.~Ruusuvuori, J.~Selinummi, H.~Huttunen, and O.~Yli-Harja.
    \newblock Computational framework for simulating fluorescence microscope images
      with cell populations.
    \newblock {\em IEEE Transactions on Medical Imaging}, 26(7):1010--1016, 2007.
    
    \bibitem{Li.2021}
    M.~Li, A.~Fang, X.~Yu, K.~Zhang, Z.~He, C.~Wang, Y.~Peng, F.~Xiao, T.~Yang,
      W.~Zhang, X.~Zheng, Q.~Zhong, X.~Liu, and Q.~Yan.
    \newblock Microbially-driven sulfur cycling microbial communities in different
      mangrove sediments.
    \newblock {\em Chemosphere}, 273:128597, 2021.
    
    \bibitem{lin2014microsoft}
    T.-Y. Lin, M.~Maire, S.~Belongie, J.~Hays, P.~Perona, D.~Ramanan,
      P.~Doll{\'a}r, and C.~L. Zitnick.
    \newblock {Microsoft COCO: Common Objects in Context}.
    \newblock In {\em {Computer Vision--ECCV 2014: 13th European Conference,
      Zurich, Switzerland, September 6-12, 2014, Proceedings, Part V 13}}, pages
      740--755. Springer, 2014.
    
    \bibitem{Liu.2021}
    B.-N. Liu, X.-T. Liu, Z.-H. Liang, and J.-H. Wang.
    \newblock Gut microbiota in obesity.
    \newblock {\em World Journal of Gastroenterology}, 27(25):3837--3850, 2021.
    
    \bibitem{Locey.2016}
    K.~J. Locey and J.~T. Lennon.
    \newblock Scaling laws predict global microbial diversity.
    \newblock {\em Proceedings of the National Academy of Sciences of the United
      States of America}, 113(21):5970--5975, 2016.
    
    \bibitem{lou2023cfpnet}
    A.~Lou, S.~Guan, and M.~Loew.
    \newblock {CFPNet-M: A Light-Weight Encoder-Decoder Based Network for
      Multimodal Biomedical Image Real-Time Segmentation}.
    \newblock {\em Computers in Biology and Medicine}, 154:106579, 2023.
    
    \bibitem{ros2}
    S.~Macenski, T.~Foote, B.~Gerkey, C.~Lalancette, and W.~Woodall.
    \newblock {Robot Operating System 2: Design, architecture, and uses in the
      wild}.
    \newblock {\em Science Robotics}, 7(66):eabm6074, 2022.
    
    \bibitem{mahecic2022event}
    D.~Mahecic, W.~L. Stepp, C.~Zhang, J.~Griffi{\'e}, M.~Weigert, and S.~Manley.
    \newblock Event-driven acquisition for content-enriched microscopy.
    \newblock {\em Nature Methods}, 19(10):1262--1267, 2022.
    
    \bibitem{doi:10.1177/2472630318784506}
    B.~Miles and P.~L. Lee.
    \newblock {Achieving Reproducibility and Closed-Loop Automation in Biological
      Experimentation with an IoT-Enabled Lab of the Future}.
    \newblock {\em SLAS TECHNOLOGY: Translating Life Sciences Innovation},
      23(5):432--439, 2018.
    
    \bibitem{fingerprint_marcel}
    M.~Molina-Moreno, M.~P. Schilling, M.~Reischl, and R.~Mikut.
    \newblock {Automated Style-Aware Selection of Annotated Pre-Training Databases
      in Biomedical Imaging}.
    \newblock In {\em {2023 IEEE 20th International Symposium on Biomedical Imaging
      (ISBI)}}, pages 1--5, 2023.
    
    \bibitem{Nduko.2020}
    J.~M. Nduko and S.~Taguchi.
    \newblock Microbial production of biodegradable lactate-based polymers and
      oligomeric building blocks from renewable and waste resources.
    \newblock {\em {Frontiers in Bioengineering and Biotechnology}}, 8:618077,
      2020.
    
    \bibitem{D3LC00327B}
    J.~P. Neto, A.~Mota, G.~Lopes, B.~J. Coelho, J.~Frazão, A.~T. Moura,
      B.~Oliveira, B.~Sieira, J.~Fernandes, E.~Fortunato, R.~Martins, R.~Igreja,
      P.~V. Baptista, and H.~Águas.
    \newblock Open-source tool for real-time and automated analysis of
      droplet-based microfluidic.
    \newblock {\em Lab Chip}, 23:3238--3244, 2023.
    
    \bibitem{Oliphant.2019}
    K.~Oliphant and E.~Allen-Vercoe.
    \newblock Macronutrient metabolism by the human gut microbiome: major
      fermentation by-products and their impact on host health.
    \newblock {\em Microbiome}, 7(1):91, 2019.
    
    \bibitem{pedone2021cheetah}
    E.~Pedone, I.~De~Cesare, C.~G. Zamora-Chimal, D.~Haener, L.~Postiglione,
      A.~La~Regina, B.~Shannon, N.~J. Savery, C.~S. Grierson, M.~Di~Bernardo,
      et~al.
    \newblock Cheetah: a computational toolkit for cybergenetic control.
    \newblock {\em ACS Synthetic Biology}, 10(5):979--989, 2021.
    
    \bibitem{pinkard2021pycro}
    H.~Pinkard, N.~Stuurman, I.~E. Ivanov, N.~M. Anthony, W.~Ouyang, B.~Li,
      B.~Yang, M.~A. Tsuchida, B.~Chhun, G.~Zhang, et~al.
    \newblock {Pycro-Manager: open-source software for customized and reproducible
      microscope control}.
    \newblock {\em Nature Methods}, 18(3):226--228, 2021.
    
    \bibitem{pinkard2022microscopes}
    H.~Pinkard and L.~Waller.
    \newblock Microscopes are coming for your job.
    \newblock {\em Nature Methods}, 19(10):1175--1176, 2022.
    
    \bibitem{susano2021python}
    D.~M.~S. Pinto, M.~A. Phillips, N.~J. Hall, J.~Mateos-Langerak, D.~Stoychev,
      T.~S. Pinto, M.~J. Booth, I.~Davis, and I.~M. Dobbie.
    \newblock {Python-Microscope – a new open-source Python library for the
      control of microscopes}.
    \newblock {\em Journal of Cell Science}, 134, 2021.
    
    \bibitem{rahmanian2022enabling}
    F.~Rahmanian, J.~Flowers, D.~Guevarra, M.~Richter, M.~Fichtner, P.~Donnely,
      J.~M. Gregoire, and H.~S. Stein.
    \newblock Enabling modular autonomous feedback-loops in materials science
      through hierarchical experimental laboratory automation and orchestration.
    \newblock {\em Advanced Materials Interfaces}, 9(8):2101987, 2022.
    
    \bibitem{ronneberger2015u}
    O.~Ronneberger, P.~Fischer, and T.~Brox.
    \newblock {U-Net: Convolutional Networks for Biomedical Image Segmentation}.
    \newblock In {\em {Medical Image Computing and Computer-Assisted Intervention
      -- MICCAI 2015}}, pages 234--241. Springer, 2015.
    
    \bibitem{sachs2022cellsium}
    C.~C. Sachs, K.~Ruzaeva, J.~Seiffarth, W.~Wiechert, B.~Berkels, and K.~N{\"o}h.
    \newblock {CellSium: versatile cell simulator for microcolony ground truth
      generation}.
    \newblock {\em Bioinformatics Advances}, 2(1):vbac053, 2022.
    
    \bibitem{Sanchez.2017}
    S.~Sanchez and A.~L. Demain.
    \newblock Useful microbial enzymes---an introduction.
    \newblock In {\em Biotechnology of Microbial Enzymes}, pages 1--11. Elsevier,
      2017.
    
    \bibitem{scherr2022microbeseg}
    T.~Scherr, J.~Seiffarth, B.~Wollenhaupt, O.~Neumann, M.~P. Schilling,
      D.~Kohlheyer, H.~Scharr, K.~N{\"o}h, and R.~Mikut.
    \newblock {microbeSEG: A deep learning software tool with OMERO data management
      for efficient and accurate cell segmentation}.
    \newblock {\em Plos one}, 17(11):e0277601, 2022.
    
    \bibitem{schilling2022kaida}
    M.~P. Schilling, S.~Schmelzer, L.~Klinger, and M.~Reischl.
    \newblock {KaIDA: a modular tool for assisting image annotation in deep
      learning}.
    \newblock {\em Journal of Integrative Bioinformatics}, 19(4):20220018, 2022.
    
    \bibitem{schmidt2022research}
    C.~Schmidt, J.~Hanne, J.~Moore, C.~Meesters, E.~Ferrando-May,
      S.~Weidtkamp-Peters, et~al.
    \newblock {Research data management for bioimaging: the 2021 NFDI4BIOIMAGE
      community survey}.
    \newblock {\em F1000Research}, 11, 2022.
    
    \bibitem{schmidt2018cell}
    U.~Schmidt, M.~Weigert, C.~Broaddus, and G.~Myers.
    \newblock {Cell Detection with Star-Convex Polygons}.
    \newblock In {\em {Medical Image Computing and Computer-Assisted Intervention
      -- MICCAI 2018}}, pages 265--273. Springer, 2018.
    
    \bibitem{dataMicrobsTrackingMillion}
    J.~Seiffarth, L.~Blöbaum, K.~Löffler, T.~Scherr, A.~Grünberger, H.~Scharr,
      R.~Mikut, and K.~N{\"o}h.
    \newblock {Data for - Tracking one in a million: Performance of automated
      tracking on a large-scale microbial data set}.
    \newblock \url{https://doi.org/10.5281/zenodo.7260137}, 10 2022.
    
    \bibitem{trackoneinamillion}
    J.~Seiffarth, L.~Blöbaum, R.~Paul, N.~Friederich, A.~J. Yamachui~Sitcheu,
      R.~Mikut, H.~Scharr, A.~Grünberger, and K.~N{\"o}h.
    \newblock Tracking one-in-a-million: Large-scale benchmark for microbial
      single-cell tracking with experiment-aware robustness metrics.
    \newblock In {\em {European Conference on Computer Vision}}. Springer, 2024.
    
    \bibitem{obiwanMicroby_microbseg}
    J.~Seiffarth, T.~Scherr, B.~Wollenhaupt, O.~Neumann, H.~Scharr, D.~Kohlheyer,
      R.~Mikut, and K.~N{\"o}h.
    \newblock {ObiWan-Microbi: OMERO-based integrated workflow for annotating
      microbes in the cloud}.
    \newblock {\em SoftwareX}, 26:101638, 2024.
    
    \bibitem{Sender.2016}
    R.~Sender, S.~Fuchs, and R.~Milo.
    \newblock Revised estimates for the number of human and bacteria cells in the
      body.
    \newblock {\em PLoS Biology}, 14(8):e1002533, 2016.
    
    \bibitem{yamachui_sitcheu_mlops_2023}
    A.~Y. Sitcheu, N.~Friederich, S.~Baeuerle, O.~Neumann, M.~Reischl, and
      R.~Mikut.
    \newblock {MLOps for Scarce Image Data: A Use Case in Microscopic Image
      Analysis}.
    \newblock In {\em {Proceedings-33. Workshop Computational Intelligence: Berlin,
      23.-24. November 2023}}, volume~23, page 169. KIT Scientific Publishing,
      2023.
    
    \bibitem{svoboda2016mitogen}
    D.~Svoboda and V.~Ulman.
    \newblock {MitoGen: A Framework for Generating 3D Synthetic Time-Lapse
      Sequences of Cell Populations in Fluorescence Microscopy}.
    \newblock {\em IEEE Transactions on Medical Imaging}, 36:310--321, 2017.
    
    \bibitem{Sylvia.2005}
    D.~M. Sylvia, J.~J. Fuhrmann, P.~G. Hartel, and D.~A. Zuberer.
    \newblock {\em Principles and applications of soil microbiology}.
    \newblock Pearson, 2005.
    
    \bibitem{celldetection}
    E.~Upschulte, S.~Harmeling, K.~Amunts, and T.~Dickscheid.
    \newblock {Contour proposal networks for biomedical instance segmentation}.
    \newblock {\em Medical Image Analysis}, 77:102371, 2022.
    
    \bibitem{wilkinson2016fair}
    M.~D. Wilkinson, M.~Dumontier, I.~J. Aalbersberg, G.~Appleton, M.~Axton,
      A.~Baak, N.~Blomberg, J.-W. Boiten, L.~B. da~Silva~Santos, P.~E. Bourne,
      et~al.
    \newblock {The FAIR Guiding Principles for scientific data management and
      stewardship}.
    \newblock {\em Scientific Data}, 3(1):1--9, 2016.
    
    \bibitem{xiao2024florence}
    B.~Xiao, H.~Wu, W.~Xu, X.~Dai, H.~Hu, Y.~Lu, M.~Zeng, C.~Liu, and L.~Yuan.
    \newblock Florence-2: Advancing a unified representation for a variety of
      vision tasks.
    \newblock In {\em {Proceedings of the IEEE/CVF Conference on Computer Vision
      and Pattern Recognition}}, pages 4818--4829, 2024.
\end{thebibliography}
\end{document}